\documentclass[prl,
aps,
,twocolumn
,superscriptaddress,
]{revtex4}
\usepackage{graphicx}

\newcommand{\ket}[1]{| #1 \rangle}

\begin{document}

\title{Boson Sampling from Gaussian States}

\newcommand{\cqct}{Centre for Quantum Computation and Communication
	Technology, School of Mathematics and Physics, University of
	Queensland, St Lucia, Queensland 4072, Australia}
\newcommand{\bristol}{Centre for Quantum Photonics, H. H. Wills Physics
	Laboratory \& Department of Electrical and Electronic Engineering,
	University of Bristol, BS8 1UB, UK}
\newcommand{\impc}{Optics Section, Blackett Laboratory, Imperial College
	London, London SW7 2AZ, United Kingdom}

\author{A. P. Lund}
\affiliation{\cqct}
\author{A. Laing}
\affiliation{\bristol}
\author{S. Rahimi-Keshari}
\affiliation{\cqct}
\author{T. Rudolph}
\affiliation{\impc}
\author{J. L. O'Brien}
\affiliation{\bristol}
\author{T. C. Ralph}
\affiliation{\cqct}
\date{\today}

\begin{abstract}

We pose a generalized Boson Sampling problem. Strong evidence exists that such
a problem becomes intractable on a classical computer as a function of the
number of Bosons. We describe a quantum optical processor that can solve this
problem efficiently based on Gaussian input states, a linear optical network
and non-adaptive photon counting measurements. All the elements required to
build such a processor currently exist. The demonstration of such a device
would provide the first empirical evidence that quantum computers can indeed
outperform classical computers and could lead to applications.

\end{abstract}


\maketitle

{\it Introduction}: Quantum computers are expected to be able to outperform
their classical counterparts for a number of different calculations
\cite{NIE00}. However, there is a large disparity between the number of quantum
bits (qubits) that can currently be coherently controlled ($\approx 10$) and
the number required for a calculation such as prime factoring on a scale that
would challenge classical computers ($\approx 10^6$). As a result there is
considerable interest in non-universal quantum computers that can solve
specific problems that are intractable to classical computation, but might
require significantly less overheads. Such devices could for the first time
experimentally demonstrate the power of quantum computing over classical
computers and might lead to technologically significant applications.

An example of a computational problem that can be solved efficiently by a
particularly simple quantum processor, but which is nonetheless believed to be
hard for classical computation, is Boson Sampling \cite{AAR11}. Consider a
passive, linear unitary transformation that takes $n$ individual Bosons and
scatters them into $m >> n$ output modes. Given a particular arrangement of
Bosons at the input, $k$, and a particular unitary $\hat U$, the problem is to
produce a fair sample of the output probability distribution, $P(l |U,k)$,
where $l$ is the arrangement of Bosons at the output. 

Aaronson and Arkipov (AA) argue that if a classical algorithm existed to
efficiently sample this distribution for a randomly chosen $\hat{U}$ then,
augmented with the ability to solve problems in the third level of the
polynomial hierarchy, it would be possible to estimate the permanent of an
arbitrary complex valued matrix -- a problem lying in the complexity class
$\# P$~\cite{VAL79} which essentially contains the full hierarchy.  Such
collapse of the polynomial hierarchy to the third level conflicts with the
widespread belief that it does not collapse to any level and therefore this
represents very strong evidence that the Boson Sampling problem is hard for a
classical computer. On the other hand the Boson Sampling problem maps directly
onto that of sampling the output photon counting distribution when $n$ single
photon states are injected into a $m \times m$ linear optical network. If such
a device was constructed then the observed output probability distribution
would be the required sample -- it would effectively be a quantum processor
that could efficiently solve the Boson Sampling problem \cite{note}. We will
refer to this device as a {\it Boson sampler}.

This observation has led to a number of proof of principle experiments in which
the basic tenets of the theory have been tested in small-scale experiments
where 3-4 photons have been injected into 5-6 mode optical networks
\cite{BRO13,SPR13,TIL13,CRE13}. Whilst the potential for scaling up the network
size for such experiments is optimistic, indeed 21 mode network experiments
with 2 input photons have already been performed \cite{PER10}, the potential
for scaling to much larger input photon numbers is more pessimistic, at least
in the short term. This is because current single photon sources are
spontaneous and so the probability for producing an $n$ photon input state
drops exponentially with $n$. Whilst deterministic single photon sources are in
development, they are only likely to provide a solution to this problem in the
medium to long term. In contrast, deterministic sources of non-classical states
with Gaussian Wigner functions (Gaussian states) of high purity{\it are}
currently available.

In this paper we ask whether a non-adaptive linear optical network which takes
a Gaussian state (as opposed to a number state) as its input, and takes photon
number counting statistics as its output, can efficiently sample distributions
that are computationally hard to sample with a classical computer. We will
refer to such a device as a {\it Gaussian Boson sampler}. We argue that for one
particular example of this type of problem the answer to this question is {\it
yes}. Specifically we show that a generalized Boson Sampling problem, in which
the task is to sample over all possible $n$ photon input distributions to the
standard Boson Sampling problem, can be efficiently solved by a Gaussian Boson
sampler. If a classical computer could solve this generalized exact Boson
sampling problem efficiently then it could, as a subset of its answer, produce
the solution to the standard Boson Sampling problem efficiently. As the latter
is believed to be impossible, so it follows that the exact generalized Boson
Sampling problem also is hard for a classical computer.  A more subtle and
physically relevant question is whether approximate Boson Sampling remains hard
in the generalised case.  We present strong evidence that this is indeed the
case.

\begin{figure}
  \centering
  \includegraphics{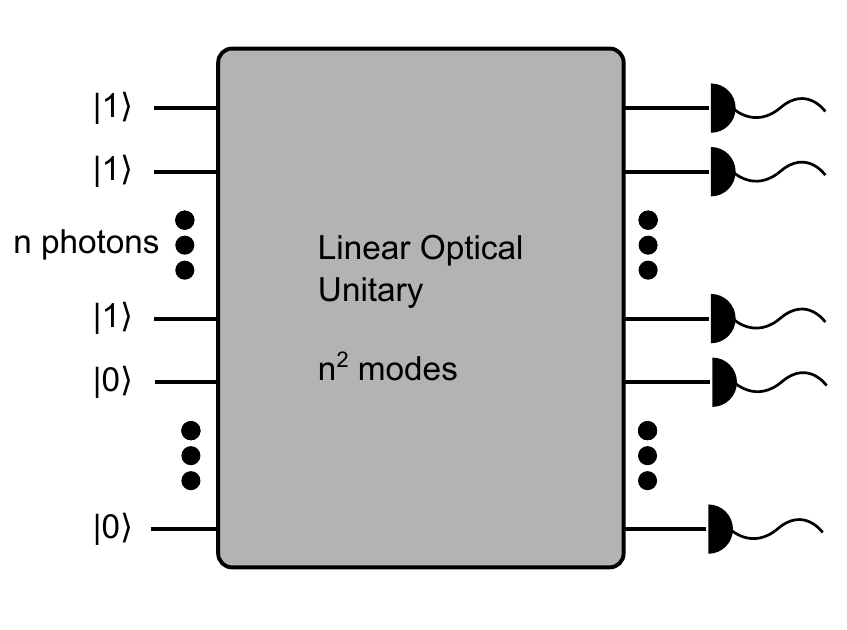}
  \caption{\label{bosonsampling} Schematic of the Boson Sampling algorithm with
  a number state input. This device efficiently samples the probability
  distribution $P(l |U,k)$.  }
\end{figure}

{\it Boson Sampling}: A schematic of a quantum optical processor which can
solve the Boson Sampling problem is shown in Fig.~\ref{bosonsampling}.
Following Ref.~\cite{AAR11} we consider the case where $m=n^2$. The initial
state is 
\begin{equation}
 \prod_{h=1}^{n^2} (\hat a_h^\dagger)^{k_h} \ket{0},
\end{equation}
where $\{k_h\}$ is a particular string of $n^2$ numbers, $n$ of which take the
value $1$ and the other $n^2-n$ take the value zero, $\hat a_h^\dagger$ is the
photon creation operator for the $h$th mode and $\ket{0}$ is the global vacuum
state of all $n^2$ modes.  After the unitary we have
\begin{equation}
\hat U \prod_{h=1}^{n^2} (\hat a_h^\dagger)^{k_h} \ket{0}.
\label{bos}
\end{equation}
AA showed that a classical sampling algorithm $C$ which takes as inputs $\hat
U$ and $\{k_h\}$ and outputs samples which are exactly those of the number
distribution represented by Eq.\ref{bos} cannot be done efficiently unless the
polynomial hierarchy of complexity classes collapses, a situation believed to
be implausible. Further AA showed that even computing samples from a
distribution close to the exact case (where the precision is an additional
input parameter) is likely to be hard for a classical computer.  This is
because the probability of particular configurations in the output is
proportional to the matrix permanent of a sub-matrix of the matrix which
linearly transforms the creation operators.  AA then used results about the
computational complexity of estimating matrix permanents to show that when one
assumes that the sampling can be done efficiently then the polynomial hierarchy
collapse occurs.  Physically, a device could be built and would efficiently
produce samples from the distribution. The results of AA suggest that such a
device has powers which are outside the classical polynomial hierarchy.  

{\it Two-mode Squeezing}: Quantum information experiments involving only a few
photons are generally performed using spontaneous parametric down-conversion.
This process involves coherently converting photons from a strong pump beam
into two modes (often called the signal and idler modes).  The output state of
these two modes is a two-mode squeezed vacuum of the form
\begin{equation}
\sqrt{1-\chi^2} \sum_{p=0}^\infty \chi^p \ket{p}_1 \ket{p}_2,
\end{equation}
where $\ket{p}_i = \hat a_i^{\dagger}/\sqrt{p!} \ket{0}$ is a $p$ photon number
state of the $i$th mode and $0 \leq \chi < 1$ is a parameter determining the
strength of the squeezing.  This state is a Gaussian state and is regularly
produced, to a good approximation, in many labs around the world.  As observed
by AA, it is possible to implement Boson Sampling with $n$ photons
post-selectively, using only two mode squeezed states as an input \cite{AAR11}.

\begin{figure}
  \centering
  \includegraphics{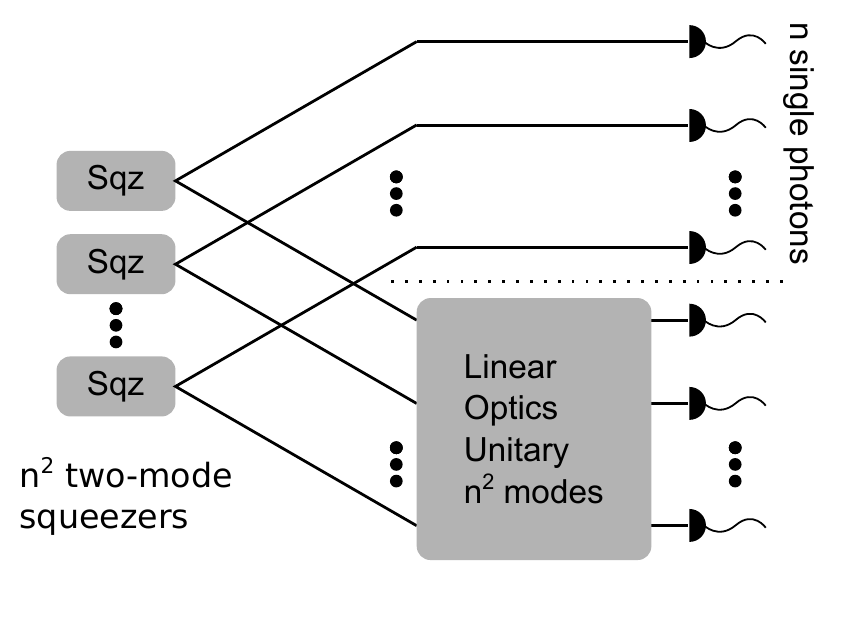}
  \caption{\label{sqzbosonsampling} Schematic of the Boson Sampling algorithm
    with squeezed state inputs and post-selection. If only a particular
    detection arrangement of $n$ photons of the upper mode are retained then
    this device can also sample the probability distribution $P(l |U,k)$, but
    not efficiently. However, if {\it all} $n$ photon arrangements are retained
    then the probability distribution $P(l, k |U)$ can be sampled efficiently.}
\end{figure}

Given a linear optical unitary on $n^2$ modes, it is possible to construct an
instance of the $n$ photon Boson Sampling problem using two-mode squeezed
vacuum states, $2n^2$ optical modes and photon counting. The configuration is
shown schematically in Fig.~\ref{sqzbosonsampling}. For each input mode in the
given unitary, one half of a two mode squeezed state is input into it.  This
therefore requires $n^2$ two mode squeezing operations.  The other half of each
state is sent directly to a photon counter.  The total state prior to detection
is
\begin{equation}
\ket{\phi}_{1,2} = U_j^{(2)} \prod_{h=1}^{n^2} \sqrt{1-\chi^2}
\sum_{p_h=0}^\infty \chi^{p} \ket{p}_{h1}\ket{p}_{h2},
\end{equation}
where the state subscript $h1$ ($h2$) refers to the set of modes that do not
(do) pass through the unitary. Consider cases when the particular arrangement
of $n$ single photons (and zero photons otherwise) described by the string
$\{k_h\}$ is counted in the set of modes $h1$. This corresponds to projecting
onto the state 
\begin{equation}
\ket{\theta}_1 =  \prod_{h=1}^{n^2} (\hat a_h^\dagger)^{k_h} \ket{0}_1.
\end{equation}
The reduced state when this particular string is measured is then 
\begin{equation}
_1\langle \theta\ket{\phi}_{1,2} =  U_j^{(2)} \prod_{h=1}^{n^2} (\hat
a_h^\dagger)^{k_h} \ket{0}_2.
\end{equation}
Comparison with Eq.\ref{bos} shows this is equivalent to an instance of the
Boson Sampling problem.  

However, the probability of detecting the particular arrangement $\{k_h\}$ is
\begin{equation}
	\label{prob:single}
  \chi^{2n} (1-\chi^2)^{n^2}.
\end{equation}
This probability adds an exponential overhead to the Boson Sampling algorithm.

For the case of exact sampling, this does not change the arguments of AA.  The
exact sampling theorem from AA uses Stockmeyer's approximate counting
algorithm~\cite{STO83} which allows for pre-factors for the probability of the
form $2^{-poly(n)}$ without changing the level of the polynomial hierarchy that
this algorithm is contained in.  That is, if the probability from the AA
Boson Sampling device was $p$, then the same probability will be found in the
distribution from our generalised Boson Sampling device as a probability
$~\chi^{2n} p$. Here we are thinking of $\chi$ being constant and hence the
pre-factor is of the form permitted and the hardness argument for exact
sampling still holds.  

Approximate Boson Sampling allows $C$ to sample from a probability
distribution, $\{q_S\}$, which is constrained in variation distance to the
exact probability distribution, $\{p_S\}$, as
\begin{equation}
\sum_S |p_S - q_S| \leq \beta ,
\label{approx}
\end{equation}
where $S$ represents configurations for Fock state detections and $\beta$ is an
input parameter.

Our aim is to show that the result from AA which shows that approximate boson
sampling implies a polynomial hierarchy collapse still holds for sampling with
squeezed state inputs.  Approximate sampling does not necessarily allow for
exponentially small scaling of the probabilities without changing its
complexity properties and so a more detailed analysis is required to show this.
The reason why this doesn't necessarily hold is because the probability of the
configuration we are interested in $S^\prime$ (i.e. the matrix permanent of a
particular sub-matrix from the unitary matrix) could be exponentially different
from that which we are expecting and yet the distribution over all
configurations might still satisfy the approximate sampling requirement of
Eq.~(\ref{approx}).  This is possible because the single probability that is of
interest could be exponentially small compared with all other probabilities.  

The key concept presented in AA is that if the sub matrix which contains the
matrix we wish to estimate the permanent of is randomly embedded in a large
random unitary matrix, then $C$ cannot know a priori which $S^\prime$
configuration is of interest.  Therefore provided Eq.~(\ref{approx}) holds and
$C$ correctly approximates most of the ${p_S}$, including the randomly chosen
$p_{S^\prime}$, with a high probability, then being able to efficiently compute
the approximate boson sampling problem enables the efficient estimation of a
complex matrix permanent and the polynomial hierarchy collapses.

We will now present two possible ways to show that approximate sampling still
implies a collapse of the PH when using the Gaussian two mode squeezed states
as resources.  The first will involve adding adaption and will not require any
additional results over what AA presented.  The second will postulate that the
adaption is not required for approximate sampling to still be hard and we will
give evidence to support this postulate.

\begin{figure}[thb]
  \centering
  \includegraphics{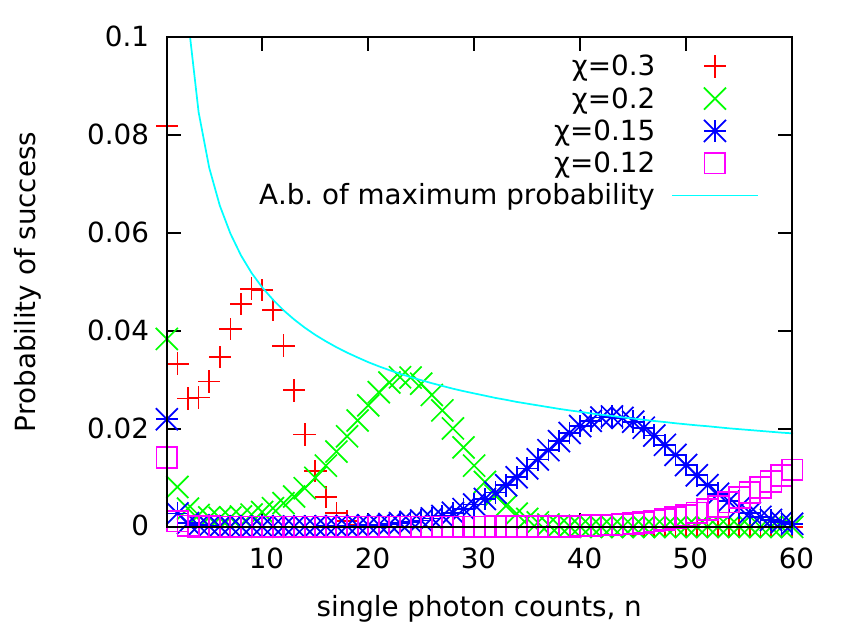}
  \caption{\label{prob}Probability of detecting $n$ single photons from $n^2$
  two mode squeezed states, irrespective of the location of the single photon
  detections.  Multiple values of $\chi$ are shown.  The asymptotic behaviour
  (a.b.) of the peak probability is shown by the solid line.  Note that the $n$
  which gives a high probability decreases with increasing $\chi$.}
\end{figure}

{\it Adding adaption}: If we consider the case where $n$ single photons are
found in the $n^2$ conditioning detectors and zero everywhere else,
irrespective of the exact location of those counts then the overall probability
is increased due to the number of ways these counts could be achieved to
\begin{equation}
{n^2 \choose n} \chi^{2n} (1-\chi^2)^{n^2}.
\end{equation}
For a given $n > 0$ this probability has a maximum when
\begin{equation}
	\label{chimax}
\chi_{max} = \frac{1}{\sqrt{n+1}}
\end{equation}
and using the Stirling approximation it is possible to show that the
asymptotic behaviour of the probability is
\begin{equation}
P(n |\chi_{max}) \sim \frac{1}{e\sqrt{2\pi}} \frac{1}{\sqrt{n-1}}.
\label{prob_eq}
\end{equation}
These results are shown graphically in Fig.~\ref{prob}. 
The location of the $n$ single photon detections is correlated with the
location of preparing a single photon state.  Hence if the input modes to the
unitary can be adapted via feed-forward so that the $n$ modes containing the
single photons are switched such that they are described by the string
$\{k_h\}$, then the Boson Sampling algorithm can be efficiently performed with
only an $O(\sqrt{n})$ overhead in running time. This is now an efficient method
for solving the Boson Sampling problem of Eq.\ref{bos}, but this has come at
the price of a requiring an adaption strategy.  Using adaption is technically
challenging but significantly less so than full fault tolerant quantum
computation. 

{\it Generalized Boson Sampling}: Here we postulate that, without adapting the
modes before the input to the unitary, this configuration will still imply a
collapse of the polynomial hierarchy in a similar way as the Boson Sampling
problem as initially described by AA.  More technically, if there is a
classical algorithm within the polynomial hierarchy for approximate sampling
from the non-adaptive configuration then the Gaussian Permanent Estimation
problem lies within the polynomial hierarchy.  The Gaussian Permanent
Estimation problem is the production of an estimate of the permanent of a
matrix whose elements are Gaussian random variables.  AA study the complexity
of this problem and show that provided the conjectures they make hold true,
then having the Gaussian Permanent Estimation problem within the polynomial
hierarchy implies a collapse of the polynomial hierarchy.  

Consider the case of measuring $n$ single photons in the heralding modes as
mentioned above in the adaptive case.  The probability of this event scales as
$1/\sqrt{n}$ (see Eq.\ref{prob}) and the exact location of the photons is
i.i.d.  In the output modes is a photon number distribution of $n$ photons
which, conditional on a particular result, is an instance of the Boson Sampling
problem.  Therefore this configuration randomly samples from ${n^2 \choose n}$
instances of the Boson Sampling problem, where both the input and output are
known. 

If the unitary matrix is chosen randomly, just as described by AA, but is
square and the rows and columns of the unitary which encode the matrix
permanent of the original problem are randomly chosen and i.i.d. then there is
no preferred event within this subspace that could be used to corrupt the
probability we are interested in whilst maintaining the conditions for
approximate sampling.  Recall that the subspace of interest is post-selected
from the entire ensemble, but is only polynomially reduced in size.  The random
matrix would still be a Harr-random unitary as presented in AA as all that is
needed to build up the rectangular matrix to a square matrix is to add in extra
random columns that are orthogonal to the columns that encode the problem.  The
approximate counting using Stockmeyer's algorithm would also proceed in the
same way as AA.  But the probability that is to be estimated would be
multiplied by the exponential pre-factor from Eq.~(\ref{prob:single}).  Note
here that we have to choose $\chi$ as per Eq.~(\ref{chimax}) to achieve the
polynomially decreasing size of the $n$ single photon counts over all
configurations.  However, this decrease does not reduce the pre-factor on the
individual probabilities fast enough for the complexity of approximate counting
to change.  The pre-factor is bounded below by $2^{-n^2}$ which as explained
above is permitted for Stockmeyer's approximate counting algorithm.  Whilst
this description does not constitute a rigorous proof and such a proof is
beyond the scope of this paper, it does strongly suggest that such a proof,
that draws from the results of AA, would be possible.

It is interesting to note that if the heralding information is discarded, then
the output configurations are almost trivial to compute.  Tracing out one mode
of the two mode squeezed state results in a thermal state with a temperature
that depends on the amount of squeezing $\chi$.  If all two mode squeezed state
sources have an equal squeezing then the reduced density matrix at the input to
the linear optical unitary is a $n^2$ mode thermal state.  The covariance
matrix for this state is proportional to the identity.  The linear optical
unitary maps covariance matrices under a symplectic map which maps the identity
to the identity.  Hence the linear optical unitary leaves this input state
unchanged.  The photon number statistics can then be easily computed from this
state.

{\it Experimental Considerations}: Challenges in building a Gaussian Boson
sampler as described above would be: (i) constructing a large number ($n^2$) of
identical two mode squeezers; (ii) injecting them into a large, highly
connected, highly coherent and low loss optical network; and (iii) counting
photons with high efficiency. One solution is to construct a large scale
integrated optics circuit, ideally with the squeezed sources and photon
counters built in to minimize loss. Recent work in which two mode squeezed
state generation plus a simple optical network were integrated into a single
chip gives reason for optimism about this approach \cite{SIL13}. Alternatively,
time multiplexed squeezed sources might offer a compact solution (see for
example \cite{MEN10}).

A major consideration for experiments will be errors, in particular loss. It is
not known if active error correction is possible given the limitations on the
allowed operations. Passive error correction against loss is possible for the
Boson sampler by simply postselecting for the desired photon number $n$
\cite{ROH12}. A similar strategy will work for the Gaussian Boson sampler
provided the losses on the directly detected modes are much smaller than those
undergoing the unitary transformation (a not unreasonable requirement given
the extra optics needed to implement the unitary would add loss).  Error
correction via postselection does not scale efficiently with the problem, going
as $\eta^n$, where $\eta$ is the average efficiency of each optical path. The
question is then whether it is possible to reach problems of an interesting
size before the efficiency requirements become too severe. We estimate that for
$n=20$ and average path transmissions of $\eta_1 = 0.99$ and $\eta_2 = 0.9$ the
postselection efficiency would be $\approx 8\%$. If we combine this with the
optimal probability of success from Fig.~\ref{prob} we conclude that $\approx
1/4 \%$ of experimental runs would produce a sample point. Given MHz repetition
rates this would still lead to thousands of samples per second. We note that
Ref. \cite{ROH12} also provided evidence that the lossy Boson Sampling problem
might itself be computationally hard. This could reduce the efficiency
requirements needed to sample a hard distribution.

{\it Conclusion}:  We have shown that a Gaussian Boson sampler -- a quantum
optical processor comprised of 2-mode squeezed state inputs, a non-adaptive
linear optical network and photon counting -- can solve a generalized Boson
Sampling problem efficiently. We have presented strong evidence that the
generalised Boson sampling problem is computationally hard for a classical
computer.  We believe this device may represent the best short term possibility
for experimentally demonstrating a quantum processor that can perform
calculations that would challenge the abilities of the best classical
processors. This represents a specific example of a unitary with Gaussian state
inputs and photon counting where the output computes a hard problem.  But, this
problem covers only a small set of possibilities when compared to the larger
problem of a general unitary with squeezed state inputs. This more general
problem will be the subject of future research.


\begin{thebibliography}{99}
\bibitem{note} Because of the intrinsically probabilistic nature of quantum
	mechanics, the ability of the quantum processor to sample the
	distribution does not imply that the quantum processor can solve for
	particular permanents.
\end{thebibliography}
\end{document}